\documentclass[prd,aps,twocolumn,nofootinbib,preprintnumbers,superscriptaddress,balancelastpage,longbibliography]{revtex4-2}

\usepackage{graphicx}
\usepackage{bm}
\usepackage{slashed}
\usepackage{aas_macros}
\usepackage{float}
\usepackage{multirow}
\usepackage[normalem]{ulem}
\usepackage{amsmath}
\usepackage{dsfont}
\usepackage{array} 
\usepackage[dvipsnames,svgnames]{xcolor}
\usepackage{tikz}
\usepackage{orcidlink}
\usepackage{fontawesome5}

\newcommand{\iconcolor}{MidnightBlue}
\protected\def\IconThong{\mbox{\textcolor{\iconcolor}{\faDog}}}
\protected\def\IconShyam{\mbox{\resizebox{!}{2.4ex}{\tikz{%
  \fill[\iconcolor] (-0.45,-0.35) .. controls (-1.0,-1.15) and (-1.6,-1.5) .. (-2.3,-1.72)
     .. controls (-1.6,-1.68) and (-0.95,-1.55) .. (-0.25,-1.15) -- cycle;%
  \fill[\iconcolor, rotate around={-28:(0.1,0.05)}] (0.1,0.05) ellipse (0.62 and 1.05);%
  \fill[\iconcolor] (-0.15,-0.75) circle (0.62);%
  \fill[\iconcolor, rounded corners=2pt] (-0.45,-1.78) -- (1.15,-1.78) -- (1.05,-1.55) -- (-0.1,-1.3) -- cycle;%
  \fill[\iconcolor] (0.35,0.75) -- (0.75,1.35) -- (1.0,1.15) -- (0.75,0.55) -- cycle;%
  \fill[\iconcolor, rotate around={-20:(0.95,1.35)}] (0.95,1.35) ellipse (0.45 and 0.26);%
  \fill[\iconcolor, rotate around={25:(0.68,1.78)}] (0.68,1.78) ellipse (0.09 and 0.3);%
  \fill[\iconcolor, rotate around={10:(0.85,1.8)}] (0.85,1.8) ellipse (0.09 and 0.3);%
  \draw[\iconcolor, line width=2.2pt, line cap=round] (0.62,0.35) -- (1.0,0.05);%
  \fill[white] (1.0,1.42) circle (0.06);%
}}}}
\protected\def\IconCarlos{\mbox{\textcolor{\iconcolor}{\faCat}}}
\protected\def\IconPedro{\mbox{\textcolor{\iconcolor}{\faDragon}}}
\protected\def\orcid#1{\orcidlink{#1}}

\makeatletter
\def\@fnsymbol#1{\ifcase#1\or\IconThong\or\IconShyam\or\IconCarlos\or\IconPedro\else\@ctrerr\fi}
\makeatother

\newcommand{\be}{\begin{equation}}
\newcommand{\ee}{\end{equation}}
\newcommand{\bea}{\begin{eqnarray}}
\newcommand{\eea}{\end{eqnarray}}

\usepackage{hyperref}
\hypersetup{
     colorlinks = true,
     citecolor  = magenta,
     urlcolor   = magenta,
     linkcolor  = magenta
}

\begin{document}

\title{Probing Dark Matter-Electron Scattering with the Central Molecular Zone}

\author{Thong T.Q. Nguyen\,\orcid{0000-0002-8460-0219}}
\thanks{\href{mailto:thong.nguyen@fysik.su.se}{thong.nguyen@fysik.su.se}}
\affiliation{Stockholm University and The Oskar Klein Centre for Cosmoparticle Physics, Alba Nova, 10691 Stockholm, Sweden}

\author{Shyam Balaji\,\orcid{0000-0002-5364-2109}}
\thanks{\href{mailto:shyam.balaji@nyu.edu}{shyam.balaji@nyu.edu}}
\affiliation{Physics Department, King’s College London, Strand, London, WC2R 2LS, United Kingdom}
\affiliation{New York University Abu Dhabi, PO Box 129188, Saadiyat Island, Abu Dhabi, UAE}

\author{Carlos Blanco\,\orcid{0000-0001-8971-834X}}
\thanks{\href{mailto:carlosblanco@psu.edu}{carlosblanco@psu.edu}}
\affiliation{Institute for Gravitation and the Cosmos, The Pennsylvania State University, University Park, PA 16802, USA}
\affiliation{Department of Physics, Princeton University, Princeton, NJ 08544, USA}
\affiliation{Stockholm University and The Oskar Klein Centre for Cosmoparticle Physics, Alba Nova, 10691 Stockholm, Sweden}

\author{Pedro De la Torre Luque\,\orcid{0000-0002-4150-2539}}
\thanks{\href{mailto:pedro.delatorre@uam.es}{pedro.delatorre@uam.es}}
\affiliation{Departamento de F\'{\i}sica Te\'orica, M-15, Universidad Aut\'onoma de Madrid, E-28049 Madrid, Spain}
\affiliation{Instituto de F\'{\i}sica Te\'orica UAM-CSIC, Universidad Aut\'onoma de Madrid, C/ Nicol\'as Cabrera, 13-15, 28049 Madrid, Spain}

\begin{abstract}
\noindent The Central Molecular Zone, a dense and gas-rich region surrounding the Galactic Center, hosts numerous dense molecular gas cores that provide unique environments to probe dark matter interactions. Among these, the IRS~13E3 cluster is one of the closest to the Galactic Center and one of the most extensively studied. We show that dark matter-electron scattering can substantially ionize molecular hydrogen (H$_2$) in the IRS~13E3 core, contributing to the ionization rate in conjunction with standard cosmic rays. Using the ionization rate inferred from H$_3^+$ observations of the central parsecs, we derive constraints on heavy and ultralight dark photon mediators, a heavy leptophilic mediator, and electric dipole moment interactions in the strongly interacting regime, where terrestrial searches are limited by the overburden problem. Our results highlight the potential of dense molecular cores near the Galactic Center as promising targets for constraining dark matter interactions with future observations.
\end{abstract}

\maketitle

\section{Introduction}
\label{sec:intro}

Understanding the interactions between dark matter (DM) and the Standard Model (SM) remains a central objective of particle physics and astronomy~\cite{Bertone:2016nfn, Bertone:2018krk, Bertone:2004pz}. As a joint effort of the physics community, direct detection experiments have achieved remarkable sensitivity to weak DM interactions with nucleons and electrons~\cite{Essig:2022dfa, Hochberg:2022apz, Akerib:2022ort, DarkSide-50:2022qzh}. From large underground detectors that search for nucleon and electron recoils from DM interactions~\cite{XENON:2025vwd, LZ:2024zvo, PandaX:2024qfu, XENON:2019gfn}, to the small scale detectors with novel targets for DM scattering~\cite{QROCODILE:2024nqm, Schwemmbauer:2025evp, Blanco:2019lrf, Blanco:2021hlm, Blanco:2022pkt, Matas:2026brk, Abbamonte:2025guf, Hochberg:2017wce, Essig:2019xkx, Tiffenberg:2017aac, Derenzo:2016fse, Essig:2015cda, Griffin:2020lgd, Griffin:2018bjn, Cook:2024cgm, DAMIC-M:2025luv, SENSEI:2020dpa}, direct detection has covered a wide mass range from MeV to GeV DM masses and now closely approaches the neutrino fog~\cite{XENON:2024ijk, OHare:2021utq, OHare:2020lva, Akerib:2022ort, Carew:2023qrj, PandaX:2024muv}.

\begin{figure}[t]
    \centering
    \includegraphics[width=1\columnwidth]{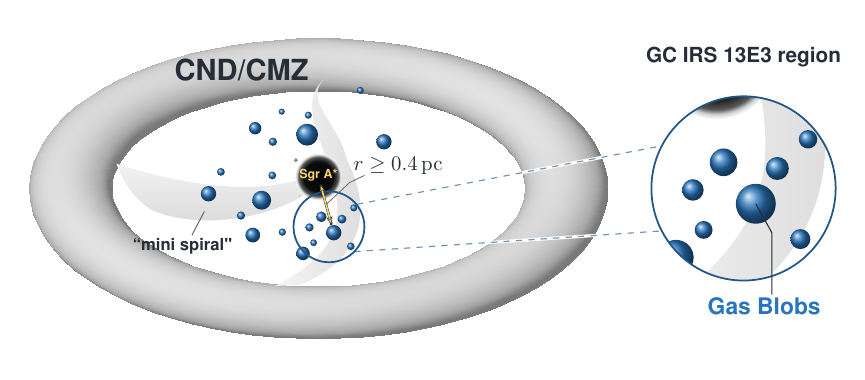}
    \caption{Schematic view of the IRS~13E3 cluster in the Galactic Center, located interior to the Circumnuclear Disk (CND) at a distance of $r \geq 0.4\,\mathrm{pc}$ from Sgr~A$^{\star}$.}
    \label{fig:CMZfig}
\end{figure}

However, these terrestrial searches face intrinsic limitations in the regime of large interaction cross sections~\cite{Starkman:1990nj}. For strongly interacting massive particle (SIMP) DM, elastic scattering with the Earth's atmosphere and overburden can attenuate the incoming flux, preventing DM particles from reaching underground detectors. This attenuation also affects other strategies that involve astrophysical phenomena, such as boosted DM by CR scatterings~\cite{Cappiello:2019qsw, Cappiello:2024acu, Alam:2025qbq, Bringmann:2018cvk}, or solar reflection~\cite{Emken:2017hnp, Emken:2021lgc, Emken:2024nox}, where the boosted DM particles still need to travel to underground detectors. 
Some cosmological probes can disfavor large DM cross sections, but these constraints weaken when the interacting DM is only a small fraction of the total DM abundance~\cite{Gluscevic:2017ywp, Boddy:2018kfv, Boddy:2018wzy, Nadler:2019zrb, Maamari:2020aqz, Nguyen:2021cnb, Boddy:2022tyt, Das:2025bnr, Ghosh:2026juc}. Therefore, the overburden problem has become an important topic of study in the DM physics community~\cite{Mahdawi:2017utm, Cappiello:2023hza, Emken:2018run, McKeen:2022poo, Lantero-Barreda:2025sgy}, as it determines the sensitivity of direct-detection experiments in the high-cross section regime and, consequently, the remaining open parameter space.

Fortunately, nature provides us with a wealth of astrophysical systems that can probe different aspects of DM-SM interactions, offering complementary probes that may already provide hints of such interactions. One of the most interesting classes of objects is molecular clouds, given their low temperatures and relatively quiescent environments. Inside these clouds, DM particles can inject additional energy through their annihilation or decay into SM particles~\cite{DelaTorreLuque:2025zjt, Leane:2025ohu, Perez:2026cqi, DelaTorreLuque:2024fcc, Balaji:2025mzv}, or through their scattering with electrons and protons in hydrogen~\cite{Prabhu:2022dtm, Blanco:2023bgz}, potentially leading to efficient ionization of molecular hydrogen (H$_2$) and contributing to the overall ionization rate, which can be probed with current observations. The idea has been applied to the well-studied molecular clouds such as L1551~\cite{Prabhu:2022dtm, Blanco:2023bgz}, HD~206165, DRAGON, G1.4-1.8+87, G357.8-4.7-55~\cite{DelaTorreLuque:2025zjt, Perez:2026cqi}.

Motivated by the potential of molecular clouds as probes of DM interactions, we focus on the Central Molecular Zone (CMZ), a gas-rich region extending to $\sim200$~pc from the Galactic Center (GC) and hosting the highest concentration of molecular hydrogen in the Galaxy~\cite{Morris:1996th, Henshaw:2022rsu}. The CMZ offers a particularly interesting environment for probing DM interactions due to the simultaneous presence of a large reservoir of molecular gas and an enhanced DM density, providing an advantage over previously studied molecular clouds. For DM that couples predominantly to electrons, DM-electron scattering can ionize H$_2$~\cite{Prabhu:2022dtm}, producing an additional contribution to the ionization rate beyond that expected from standard CR processes~\cite{Ravikularaman:2024umo, Recchia:2019opr, Phan:2018owc}. This excess ionization therefore provides a sensitive probe of DM-electron interactions.

Since the ionization signal from DM-electron scattering scales linearly with the local DM density, regions close to the GC can have ionization rates many orders of magnitude larger than those in local molecular clouds for the same cross section. At the same time, observations of H$_3^+$ have characterized the ionization state of molecular gas in the CMZ and provide an empirical measure of the total ionization rate~\cite{CRAstrochemistry, Gabici:2022nac, CRIonization_Disk, CloudIon}. The CMZ therefore acts as a natural calorimeter for DM interactions. We can constrain the DM-induced ionization directly from observations, and the only model requirement is that DM scatters with electrons.

In this paper, we show that the ionization rate in the CMZ provides a sensitive probe of DM-electron interactions. We focus on the molecular gas of the GC~IRS~13E3 cluster, located at $r \gtrsim 0.4$~pc from Sgr~A$^{\star}$~\cite{Tsuboi1}, as shown in Fig.~\ref{fig:CMZfig}, and compare the DM-induced ionization with the ionization rate measured in the CMZ through H$_3^+$. We consider heavy and ultralight dark photon mediators, a leptophilic contact interaction, and an EDM interaction, and derive constraints in the strongly interacting regime, including the case where DM is a subdominant fraction of the total DM abundance. We compare our CMZ constraints with existing astrophysical and terrestrial probes and show that this approach complements them.

This paper is organized as follows. In Sec.~\ref{sec:DMmodel}, we describe the DM models and their interactions with electrons. In Sec.~\ref{sec:Ionization_rate}, we compute the ionization rate of H$_2$ from DM-electron scattering. In Sec.~\ref{sec:observation}, we discuss current observations of the CMZ ionization rate, focusing on the region closest to Sgr~A$^{\star}$. In Sec.~\ref{sec:result}, we present our constraints and discuss their implications. We conclude in Sec.~\ref{sec:conclusion}. We also study how different DM density profiles affect our constraints in Appendix~\ref{appen:uncertainty}, and how DM scattering in the cloud attenuates the DM flux in Appendix~\ref{appen:atten}.

\section{Dark Matter models}
\label{sec:DMmodel}

We consider one of the simplest extensions of the SM, a Dirac fermion DM particle $\chi$ that scatters with electrons through four different interactions: a heavy dark photon, a heavy leptophilic mediator, an ultralight dark photon, and via an electric dipole moment (EDM).

Notably, a Dirac fermion can also annihilate into SM particles through the same interactions, such as \mbox{$\chi\bar{\chi} \rightarrow e^+e^-$} by crossing symmetry~\cite{Profumo:2013hqa}, and also into other SM particles. For symmetric DM, indirect detection constrains these couplings and therefore the scattering cross section~\cite{Cirelli:2026omb, Koechler:2026pnz, Nguyen:2025ygc, Berlin:2024lwe, Nguyen:2026apa}. Our constraints do not depend on annihilation, since molecular cloud ionization requires only DM-electron scattering, so they apply to both symmetric and asymmetric DM. For asymmetric DM~\cite{Kaplan:2009ag, Zurek:2013wia, Petraki:2013wwa, Das:2024tmx, Murase:2016nwx}, the large interaction strength removes the symmetric component in the early Universe~\cite{Lin:2011gj}, and annihilation signals vanish. We also allow $\chi$ to be a subdominant DM component and define its fraction of the total DM density as \mbox{$f_\chi = \rho_\chi/\rho_{\rm DM}$}.

In the dark photon model, $\chi$ is charged under a dark $U(1)_{D}$ gauge symmetry, and the dark photon $A^{\prime}_{\mu}$ mixes kinetically with the SM photon. The Lagrangian is~\cite{Fabbrichesi:2020wbt}
\begin{equation}
\begin{split}
    \mathcal{L} \supset  \bar{\chi} &\left(i\gamma^\mu D_\mu - m_\chi\right)\chi
- \frac{1}{4}F'_{\mu\nu}F'^{\mu\nu} \\
& + \frac{1}{2}m_{A'}^2 A'_\mu A'^\mu - \frac{\varepsilon}{2}F_{\mu\nu}F'^{\mu\nu},
\label{eq:L_DP}
\end{split}
\end{equation}
where $D_\mu = \partial_\mu - i g_D A'_\mu$, $g_D$ is the dark gauge coupling with $\alpha_D = g_D^2/(4\pi)$, $m_{A'}$ is the dark photon mass, $\varepsilon$ is the kinetic mixing parameter, and $F_{\mu\nu}$ ($F'_{\mu\nu}$) is the photon (dark photon) field strength. Through the mixing term, the dark photon acquires a coupling to electrons with strength $\varepsilon e$.

The differential DM-electron cross section for this model can be written as~\cite{Lin:2022hnt}
\begin{equation}
\frac{d\sigma}{dq^2} = \frac{\bar{\sigma}_e}{4\mu_{\chi e}^2 v^2}\left|F_{\rm DM}(q)\right|^2,
\label{eq:dsigma}
\end{equation}
where $q$ is the momentum transfer, $v$ is the DM velocity, $\mu_{\chi e}$ is the DM-electron reduced mass, $\bar{\sigma}_e$ is the reference cross section, and $F_{\rm DM}(q)$ is the DM form factor, normalized so that $F_{\rm DM}(\alpha m_e) = 1$, where $\alpha$ is the fine structure constant. For the dark photon model
\begin{equation}
\bar{\sigma}_e = \frac{16\pi\alpha\alpha_D\varepsilon^2\mu_{\chi e}^2}{\left(m_{A'}^2 + \alpha^2 m_e^2\right)^2},
\qquad
F_{\rm DM}(q) = \frac{\alpha^2 m_e^2 + m_{A'}^2}{q^2 + m_{A'}^2}.
\label{eq:sigma_DP}
\end{equation}
When $m_{A'}$ is much larger than the momentum transfer, the mediator is heavy and $F_{\rm DM}(q) = 1$. When \mbox{$m_{A'} \ll \alpha m_e$}, the mediator is ultralight and \mbox{$F_{\rm DM}(q) = (\alpha m_e/q)^2$}.

Our results also apply to the case in which DM is leptophilic~\cite{Fox:2008kb} and interacts with leptons through a heavy mediator, where direct detection searches suffer less from the overburden effect. For momentum transfers well below the mediator mass, the interaction is a contact operator~\cite{Fan:2010gt, Fitzpatrick:2012ix, Catena:2019gfa},
\begin{equation}
    \mathcal{L}\supset \frac{1}{\Lambda^{2}}\,\bar{\chi}\chi\,\bar{\ell}\ell,
\end{equation}
which reduces to the operator $\mathcal{O}_{1}=\mathds{1}_{\chi}\mathds{1}_{\ell}$ in the nonrelativistic limit. The cross section takes the form of Eq.~\eqref{eq:dsigma} with $\bar{\sigma}_{e}=\mu_{\chi e}^{2}/(\pi\Lambda^{4})$ and $F_{\rm DM}(q)=1$. The form factor is the same as for the heavy dark photon, so the CMZ constraint on $\bar{\sigma}_{e}$ derived for the heavy dark photon also applies to this case. Unlike the dark photon, this interaction couples DM to protons only through loops, with a cross section suppressed by powers of $\alpha/\pi$~\cite{Kopp:2009et}.

In the EDM model, $\chi$ couples directly to the photon through the mass-dimension 5 operator~\cite{Sigurdson:2004zp}
\begin{equation}
\mathcal{L} \supset -\frac{i}{2}\, d_\chi\, \bar{\chi}\sigma^{\mu\nu}\gamma^5\chi F_{\mu\nu},
\label{eq:L_EDM}
\end{equation}
where $\sigma^{\mu\nu} = \frac{i}{2}[\gamma^\mu,\gamma^\nu]$ and the EDM coefficient $d_\chi$ has mass dimension $-1$. The cross section takes the form of Eq.~\eqref{eq:dsigma} with
\begin{equation}
\bar{\sigma}_e = \frac{4 d_\chi^2 \mu_{\chi e}^2}{\alpha m_e^2},
\qquad
F_{\rm DM}(q) = \frac{\alpha m_e}{q}.
\label{eq:sigma_EDM}
\end{equation}

The EDM interaction in Eq.~\eqref{eq:L_EDM} is an effective operator. Its coefficient has mass dimension $-1$, so new states at a scale $\Lambda$ generate it, with \mbox{$d_{\chi}\Lambda \sim 1$}. The description is self-consistent only if \mbox{$\Lambda \gtrsim m_{\chi}$}, so that the states that generate the dipole cannot be produced in the scattering. This condition gives~\cite{Sigurdson:2004zp}
\begin{equation}
    d_\chi m_\chi \lesssim 1 .
    \label{eq:EFT}
\end{equation}
Notably, partial-wave unitarity of the annihilation into two photons leads to a weaker bound \mbox{$d_\chi m_\chi \lesssim 3$}~\cite{Sigurdson:2004zp}. Therefore, we adopt the more restrictive condition of Eq.~\eqref{eq:EFT}. Above these values, a large $d_{\chi}$ requires a strongly coupled or composite dark sector~\cite{Cline:2021itd, Kribs:2016cew, Cacciapaglia:2020kgq, Abe:2024mwa}.

The four interactions therefore differ in the form factor, $F_{\rm DM}(q) = (\alpha m_e/q)^n$, with $n = 0$ for the heavy mediators, $n = 2$ for the ultralight dark photon mediator, and $n = 1$ for the EDM~\cite{Chang:2019xva, Essig:2022dfa}, and in whether DM also couples to protons.

Notably, the dark photon and EDM interactions also enables DM to scatter with protons. The dark photon mixes with the SM photon, so it couples to every charged particle with a strength proportional to its charge. Since the electron and the proton carry charges of equal size, the dark photon couples to both with strength $\varepsilon e$. The EDM interacts with the electric field of any charged particle, so it also acts on protons. This proton coupling has two consequences in the strongly interacting regime. First, nuclear scattering in the Earth dominates the attenuation of the DM flux and sets the upper reach of underground experiments~\cite{Emken:2019tni}. Second, DM scatters with protons in the cloud, with \mbox{$\bar{\sigma}_p = (\mu_{\chi p}/\mu_{\chi e})^2\bar{\sigma}_e \sim 10^4\,\bar{\sigma}_e$} at \mbox{$m_\chi = 50$~MeV.} These nuclear recoils can ionize H$_2$ through the molecular Migdal effect, in which the electrons cannot follow the sudden kick of the nucleus and one of them is ejected~\cite{Blanco:2022pkt, Blanco:2023bgz}. The same scattering also slows the DM before it reaches the inner part of the cloud and could therefore reduce the ionization from DM-electron scattering there. For the leptophilic interaction, DM scatters only with electrons, so it loses energy in the cloud much more slowly and the attenuation sets in at larger cross sections. In this work, we consider only DM-electron scattering as the source of ionization. For the dark photon and EDM interactions, our constraints are reliable well below the attenuation thresholds shown in Fig.~\ref{fig:Uncertainty}. We discuss how proton scattering attenuates the incoming DM flux, and how this affects our constraints, in Appendix~\ref{appen:atten}. We leave a full treatment of the proton channel, including Migdal ionization, to future work.

\section{CMZ ionization by DM-electron scattering}
\label{sec:Ionization_rate}

In molecular clouds, CRs, either directly produced from sources within the cloud or penetrating from the interstellar medium, ionize molecular hydrogen directly
\begin{equation}
{\rm CR} + {\rm H}_2 \rightarrow {\rm CR} + {\rm H}_2^+ + e^- .
\label{eq:CR_H2}
\end{equation}
The ${\rm H}_2^+$ ion subsequently reacts with ${\rm H}_2$ to form ${\rm H}_3^+$, whose abundance can be inferred from observations of its ro-vibrational transition lines. The observed ${\rm H}_3^+$ abundance provides a measure of the total CR ionization rate, $\zeta_{\rm CR}$~\cite{Phan:2018owc, Phan:2024fpo, Ravikularaman:2024umo, Recchia:2019opr}. However, chemical interactions with other species in the gas can alter the equilibrium abundance of ${\rm H}_3^+$ and must therefore be accounted for when inferring $\zeta_{\rm CR}$ from observations. DM that scatters with electrons ionizes ${\rm H}_2$ in the same way,
\begin{equation}
\chi + {\rm H}_2 \rightarrow \chi + {\rm H}_2^+ + e^- ,
\label{eq:DM_H2}
\end{equation}
and contributes to the observed ionization rate. We calculate the ionization rate induced by DM, without including the contribution from CRs, which are likely the primary source of ionization. Our calculation follows the approach of Ref.~\cite{Prabhu:2022dtm}.

In the scattering, DM transfers a momentum $q$ and an energy $\Delta E$ to an electron bound to the hydrogen molecule. The electron escapes when $\Delta E$ exceeds the ${\rm H}_{2}$ ionization energy~\cite{2009JChPh.130q4306L},
\begin{equation}
    E_{\rm ion} \simeq 15.43\text{ eV},
    \label{eq:Eion}
\end{equation}
and carries away a kinetic energy
\begin{equation}
    E_{r}=\Delta E - E_{\rm ion}.
    \label{eq:Erecoil}
\end{equation}
The momentum of the ejected electron is therefore
\begin{equation}
k = \sqrt{2 m_e E_r} .
\label{eq:k}
\end{equation}
The binding energy enters only through $\Delta E$, which sets the minimum DM speed for the process
\begin{equation}
v_{\rm min} = \frac{q}{2 m_\chi} + \frac{E_r + E_{\rm ion}}{q} .
\label{eq:vmin}
\end{equation}

A DM component that makes up a fraction $f_\chi$ of the total DM density has $\rho_\chi(r) = f_\chi \times \rho_{\rm DM}(r)$, where $r$ is the distance from Sgr~A$^\star$. The differential ionization rate per ${\rm H}_2$ molecule is given by~\cite{Prabhu:2022dtm, Lin:2022hnt}
\begin{align}
\label{eq:dxi}
\frac{{\rm d}\xi_{\rm DM}(r)}{{\rm d}\ln E_r}
=& \frac{\rho_\chi(r)}{m_\chi}\frac{\bar{\sigma}_e}{8\mu_{\chi e}^2} N_{T}\\
&\times \int {\rm d}q\, q \left|F_{\rm DM}(q)\right|^2
\left|f_{\rm ion}^{{\rm H}_2}(k,q)\right|^2 \eta(v_{\rm min}, r),\nonumber
\end{align}
where $N_{T} = 2$ is the number of target electrons per hydrogen molecule, $\bar{\sigma}_e$ and $F_{\rm DM}(q)$ are defined in Sec.~\ref{sec:DMmodel}, $f_{\rm ion}^{{\rm H}_2}$ is the ${\rm H}_2$ ionization form factor, and $\eta$ is the mean inverse DM speed in the cloud frame. We integrate over all $q$ that satisfy $v_{\rm min} \leq v_e(r) + v_c(r)$, where $v_e(r)$ is the local escape speed.

\begin{figure*}[t]
    \centering
    \includegraphics[width=2\columnwidth]{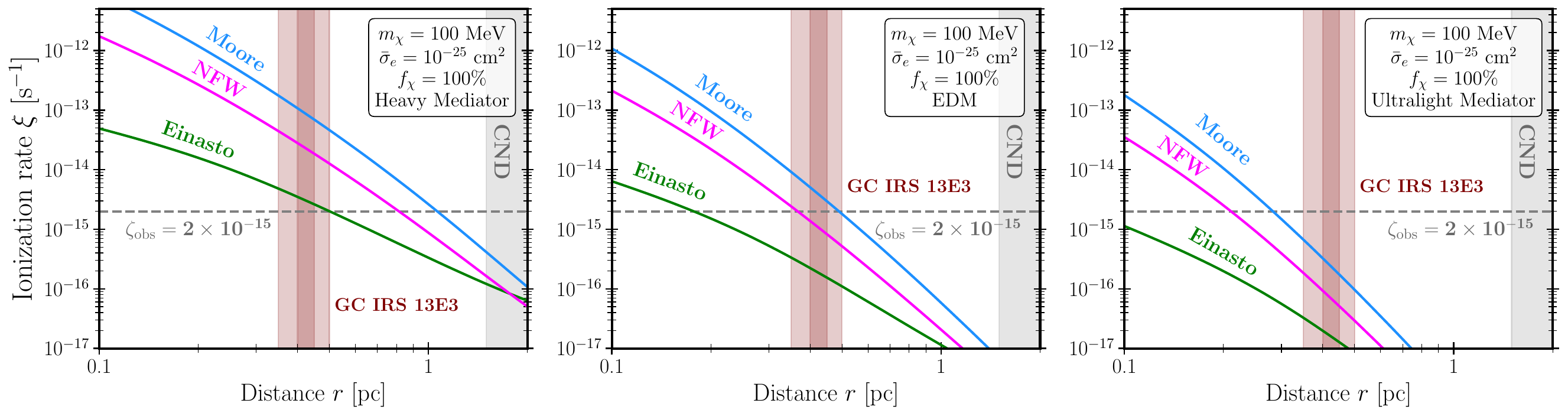}
    \caption{Ionization rate per H$_2$ molecule from DM-electron scattering as a function of distance from Sgr~A$^{\star}$, for $m_\chi = 100$~MeV and $\bar{\sigma}_e = 10^{-25}$~cm$^2$. The panels show the heavy mediator with $F_{\rm DM} = 1$ ({\bf left}), the EDM with $F_{\rm DM} \propto 1/q$ ({\bf middle}), and the ultralight mediator with $F_{\rm DM} \propto 1/q^2$ ({\bf right}), for the Einasto (green), NFW (magenta), and Moore (blue) profiles. The dashed line shows the observed ionization rate $\zeta_{\rm obs} = 2\times10^{-15}$~s$^{-1}$~\cite{Goto:2013uea, Oka:2005uq}. The gray band marks the CND, and the maroon bands mark the two radial intervals of the IRS~13E3 cloud, $[0.35, 0.45]$ and $[0.4, 0.5]$~pc.}
    \label{fig:Ionrate}
\end{figure*}

We approximate the ${\rm H}_{2}$ form factor with the hydrogen form factor~\cite{Prabhu:2022dtm},
\begin{equation}
\left|f_{\rm ion}^{{\rm H}_2}(k,q)\right|^2 \approx \frac{4.85}{2} \left|f_{\rm ion}^{\rm H}(k,q)\right|^2 .
\label{eq:fH2}
\end{equation}
We take $N_T = 2$ for the two target electrons per H$_2$ molecule and use \mbox{$|f_{\rm ion}^{{\rm H}_2}|^2 \simeq (4.85/2)\,|f_{\rm ion}^{\rm H}|^2$}, so that the combination entering Eq.~\eqref{eq:dxi} is \mbox{$N_T |f_{\rm ion}^{{\rm H}_2}|^2 \simeq 4.85\,|f_{\rm ion}^{\rm H}|^2$}, following Ref.~\cite{Prabhu:2022dtm}. The total factor of 4.85 includes both electrons of H$_2$ and the coherent enhancement from the overlap of the two atomic orbitals. It carries an uncertainty of order unity, up to a factor of about two from the normalization of the valence-bond state, which propagates linearly into the constraint on $\bar{\sigma}_e$. Eq.~\eqref{eq:dxi} therefore gives the rate per H$_2$ molecule, which we compare directly with $\zeta_{\rm obs}$. For atomic hydrogen, the ionization form factor has the closed form~\cite{Chen:2015pha, Lin:2022hnt, ARHolt_1969, DBelkic_1981, GRAVIELLE199253, Nordsieck:1954zz}
\begin{align}
\left|f_{\rm ion}^{\rm H}(k,q)\right|^2
&= \frac{512 Z^6 k^2 q^2 a_0^4 \left[(3q^2 + k^2)a_0^2 + Z^2\right]}
{3\left[(q+k)^2 a_0^2 + Z^2\right]^3 \left[(q-k)^2 a_0^2 + Z^2\right]^3} \nonumber\\
\times & \frac{\exp \left[-\dfrac{2Z}{k a_0}\arctan\left(\dfrac{2 Z k a_0}{(q^2 - k^2)a_0^2 + Z^2}\right)\right]}
{1 - e^{-2\pi Z/(k a_0)}},
\label{eq:fH}
\end{align}
where \mbox{$a_0 = 1/(\alpha m_e) \simeq 2.68\times10^{-4}~{\rm eV}^{-1}$} is the Bohr radius and $Z = 1$. The arctangent is taken on the branch $[0,\pi]$, i.e. $\pi$ is added when $(q^2-k^2)a_0^2+Z^2<0$, so that $|f_{\rm ion}^{\rm H}|^2$ is continuous across $k a_0=\sqrt{q^2a_0^2+Z^2}$.

The cloud moves with a circular velocity $v_c(r)$ relative to the halo rest frame. In the halo frame, the DM velocity $\vec{u}$ follows a Maxwell-Boltzmann distribution with dispersion $\sigma_v(r)$, truncated at the local escape speed, $|\vec{u}| < v_e(r)$. In the cloud frame, $\vec{v} = \vec{u} - \vec{v}_c$, and the DM speed follows
\begin{align}
f_{\rm cl}(v,r) = \frac{v}{\sqrt{\pi}\,\sigma_v v_c\,K(r)}
\Big[&e^{-(v - v_c)^2/\sigma_v^2} \nonumber\\
&- e^{-\min\left[(v + v_c)^2,\; v_e^2\right]/\sigma_v^2}\Big],
\label{eq:fcl}
\end{align}
for $v \leq v_e + v_c$, the largest DM speed in the cloud frame, with the normalization
\begin{equation}
K(r) = {\rm Erf}\left(\frac{v_e}{\sigma_v}\right) - \frac{2}{\sqrt{\pi}}\,\frac{v_e}{\sigma_v}\,e^{-v_e^2/\sigma_v^2}.
\label{eq:K}
\end{equation}
The mean inverse speed is then
\begin{equation}
\eta(v_{\rm min}, r) = \int_{v_{\rm min}}^{v_e + v_c} {\rm d}v\, \frac{f_{\rm cl}(v,r)}{v}.
\label{eq:eta}
\end{equation}

We compute the $\rho_{\rm DM}(r)$, $\sigma_v(r)$, $v_c(r)$, and $v_e(r)$ for Milky Way mass model~\cite{Hussein:2025xwm}, and for three different DM density profiles: Navarro-Frenk-White (NFW)~\cite{Navarro:1995iw, Navarro:1996gj}, Moore (or generalized NFW profile) with $\gamma=1.16$~\cite{Diemand:2004wh}, and Einasto~\cite{Graham:2005xx} profiles. At the sub-pc radii relevant in this study, the gravitational potential is strongly affected by Sgr~A$^\star$ and the nuclear stellar cluster, and these components are therefore included when determining the radial velocity profiles. The driving uncertainty is the DM density itself, and we quantify this dependence by presenting the limits for several representative inner-halo profiles in the Appendix~\ref{appen:uncertainty}. We use the NFW profile as our benchmark.

The total ionization rate at distance $r$ is
\begin{equation}
\xi_{\rm DM}(r) = \int_0^{E_r^{\rm max}} \frac{{\rm d}E_r}{E_r}\,\frac{{\rm d}\xi_{\rm DM}(r)}{{\rm d}\ln E_r},
\label{eq:xi_DM}
\end{equation}
where \mbox{$E_r^{\rm max} = m_\chi (v_e + v_c)^2/2 - E_{\rm ion}$} is the largest kinetic energy that DM can give to the electron. For a molecular cloud that spans from $r_1$ to $r_2$ in distance from Sgr~A$^{\star}$, we average the DM ionization rate over the cloud size \mbox{$\Delta r = r_2 - r_1$},
\begin{equation}
\bar{\xi}_{\rm DM}(m_\chi, \bar{\sigma}_e) = \frac{1}{\Delta r}\int_{r_1}^{r_2} {\rm d}r\, \xi_{\rm DM}(r).
\label{eq:xiavg}
\end{equation}
This average equals the gas-density-weighted average when the H$_2$ density is approximately uniform across the cloud. We choose $r_1$ and $r_2$ based on the distance of the GC~IRS~13E3 cluster from Sgr~A$^{\star}$, which we discuss in Sec.~\ref{sec:observation}. We exclude the parameter points where $\bar{\xi}_{\rm DM}$ exceeds the observed ionization rate $\zeta_{\rm obs}$. Since CRs and other sources also contribute to $\zeta_{\rm obs}$, constraining $\bar{\sigma}_{e}$ with this criterion is conservative. Ejected electrons with $E_r > E_{\rm ion}$ can also ionize further molecules~\cite{DelaTorreLuque:2024fcc}. We do not include this secondary ionization, which can only increase the DM-induced rate, so neglecting this effect makes our limits more conservative.

\section{The GC IRS~13E3 cluster and the CMZ ionization rate}
\label{sec:observation}

The CMZ extends to about $200$~pc from the Galactic Center and contains the highest concentration of molecular hydrogen in the Galaxy~\cite{Morris:1996th, Henshaw:2022rsu}. Toward its center, within a few parsecs of Sgr~A$^\star$, the dominant molecular feature is the Circumnuclear Disk (CND), a clumpy ring of dense molecular gas that orbits the central black hole with an inner radius of about $1.5$~pc and typical hydrogen nucleus densities $n_{\rm H} \sim 4\times10^5~{\rm cm^{-3}}$~\cite{CNDpreview1, CNDpreview2, CND_density}. Inside this ring lies a central cavity that contains ionized gas streams, including the Northern Arm, dense star clusters and gas clumps or cores.

Within this cavity, IRS~13E is a compact association of massive stars and concentrations of gas and dust, including the prominent IRS~13E3 core~\cite{Fritz:2010cr}. Gas streams on eccentric orbits can carry dense material toward Sgr~A$^\star$, allowing molecular clumps to penetrate the cavity before their molecules are destroyed by the intense radiation and CR environment~\cite{Goto:2013uea}. Their survival therefore depends on both their true distance from Sgr~A$^\star$ and their exposure time to these conditions. The Atacama Large Millimeter/submillimeter Array (ALMA)~\cite{Wootten:2009qb} can investigate this material through observations of molecular emission and hydrogen recombination lines at subarcsecond resolution~\cite{Tsuboi:2017ocm, 2019PASJ...71..105T}. These measurements resolve compact gas structures and their line-of-sight velocities, allowing the molecular and ionized components to be compared and their physical association to be assessed. Together, the gas kinematics and molecular survival provide constraints on the location of IRS~13E3 within the cavity~\cite{Tsuboi1}.

We target the molecular gas associated with the IRS~13E3 cluster in this cavity. The projected distance of IRS~13E3 from Sgr~A$^{\star}$ is only 0.13~pc~\cite{Genzel:1966zz}. Using observations with ALMA, Tsuboi et al.~\cite{Tsuboi1} detected CH$_3$OH emission in this gas. CRs destroy CH$_3$OH near Sgr~A$^{\star}$: along the Northern Arm, the CH$_3$OH emission disappears at a three-dimensional distance of about 0.4~pc, which they derived from the Keplerian orbit of the arm. Since the gas of IRS~13E3 still contains CH$_3$OH, they placed a lower limit on the three-dimensional distance of the cluster center from Sgr~A$^{\star}$, $r \gtrsim 0.4$~pc, as illustrated in Fig.~\ref{fig:CMZfig}. At this distance, the DM density is expected to be much larger than in any molecular cloud studied before, for any of the profiles considered here, by four orders of magnitude for NFW (Appendix~\ref{appen:uncertainty}).

\begin{figure*}[t]
    \centering
    \includegraphics[width=1\columnwidth]{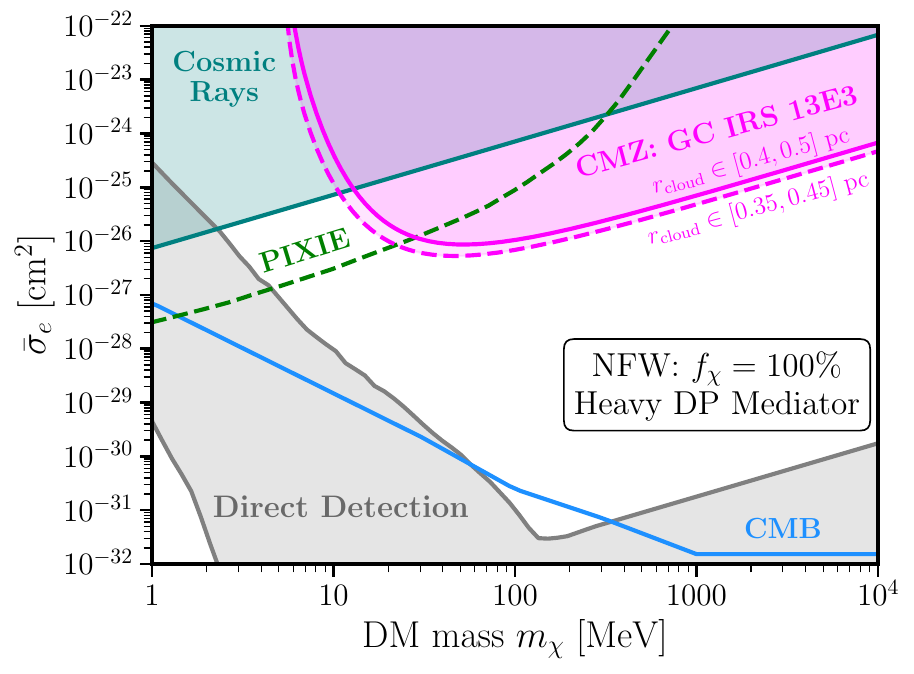}
    \hfill
    \includegraphics[width=1\columnwidth]{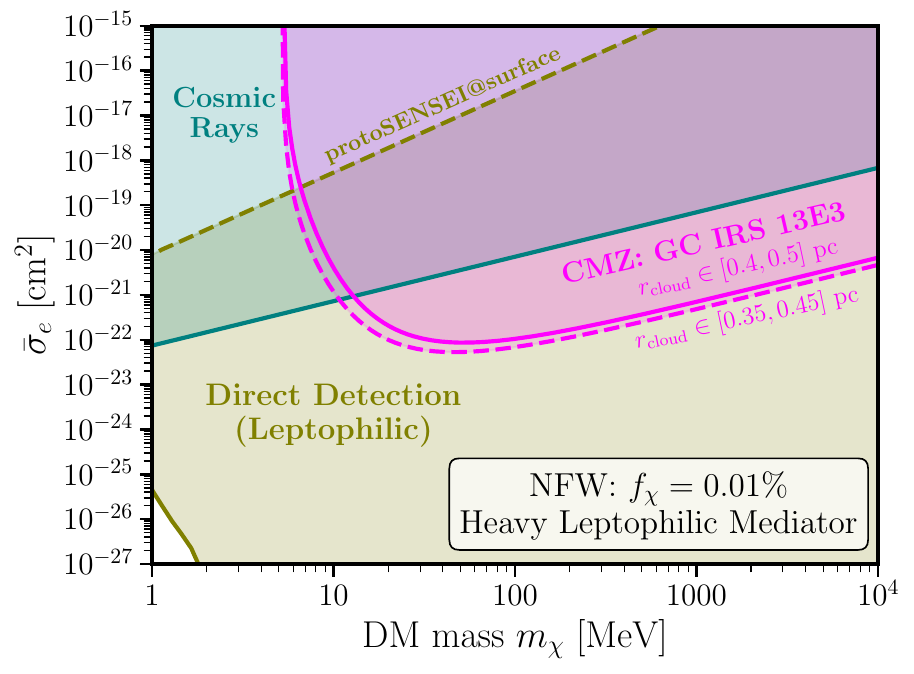}
    \caption{Constraints on the reference DM-electron cross section $\bar{\sigma}_e$ for heavy mediators, $F_{\rm DM}(q) = 1$, assuming an NFW profile. The left panel shows a heavy dark photon with $f_\chi = 100\%$, and the right panel shows a leptophilic interaction with $f_\chi = 0.01\%$. The solid and dashed magenta curves correspond to $r_{\rm cloud} \in [0.4, 0.5]$~pc and $r_{\rm cloud} \in [0.35, 0.45]$~pc, respectively; the parameter space above these curves overproduces the observed molecular ionization. We also show the regions excluded by direct detection for the heavy dark photon (gray) and for the leptophilic interaction (olive)~\cite{Emken:2019tni}, where the upper edge of the olive region is set by the protoSENSEI surface run (dashed olive)~\cite{Crisler:2018gci}, the exclusion from CR scattering above the teal curve~\cite{Cappiello:2018hsu}, the CMB bound (blue)~\cite{Xu:2018efh}, and the projected PIXIE sensitivity (dashed green)~\cite{Ali-Haimoud:2015pwa}. The CMB and PIXIE bounds do not apply at $f_\chi = 0.01\%$.}
    \label{fig:constraint_heavy}
\end{figure*}

No H$_3^+$ measurement exists for the gas associated with IRS~13E3 itself. The closest determination is that of Goto et al.~\cite{Goto:2013uea}, who derived \mbox{$\zeta_2 = 1.2\times10^{-15}$~s$^{-1}$} for the dense gas at $+50$~km\,s$^{-1}$ on the line of sight to GCIRS~3, a few tenths of a parsec from Sgr~A$^\star$ in projection, and regard this value as a lower limit. For the diffuse gas of the CMZ as a whole, Oka et al.~\cite{Oka:2005uq} find \mbox{$\zeta_2 = (2$--$7)\times10^{-15}$~s$^{-1}$}. We adopt $\zeta_{\rm obs} = 2\times10^{-15}$~s$^{-1}$ as the total ionization rate of the IRS~13E3 gas requiring $\bar{\xi}_{\rm DM}\le\zeta_{\rm obs}$. Since the CR intensity is expected to rise toward Sgr~A$^\star$, which is the argument that places the CH$_3$OH-containing gas at $r\gtrsim0.4$~pc~\cite{Tsuboi1}, this value is the low end of the measured range. Adopting the upper end of the CMZ range, $7\times10^{-15}$~s$^{-1}$, or a cap of $2\times10^{-14}$~s$^{-1}$, would weaken every bound by a factor of around 3.5 or 10 respectively, since the limits scale linearly with $\zeta_{\rm obs}$.

We calculate the ionization rate due to DM-electron scattering as a function of distance from Sgr~A$^{\star}$ for each DM density profile, as shown in Fig.~\ref{fig:Ionrate}. The rate falls by three orders of magnitude between 0.1 and 1 pc. The DM density accounts for one order of magnitude in the case of an NFW profile. The rise of the escape and orbital velocities toward Sgr~A$^\star$ accounts for the additional suppression of $\xi$ with increasing $r$, since a larger fraction of the DM is above the ionization threshold. The steep profiles, Moore and NFW, give rates around 20 and 5 times the Einasto rate at 0.4 pc. The ALMA observations achieve an angular resolution as small as \mbox{$0.41''\times 0.3''$}, corresponding to a physical resolution of approximately $0.016\times 0.012$~pc for targets in the GC region. The spatial extent of the molecular material associated with IRS~13E3 is not established by these observations. We therefore adopt a characteristic spatial scale of \mbox{$L_{\rm cloud}=\Delta r = 0.1\,\mathrm{pc}$} for the molecular material. This value is chosen as a conservative estimate of the cloud size, comparable to the characteristic sizes of compact molecular structures observed with ALMA in the central cavity~\cite{Goicoechea2018}. We then calculate the average ionization rate using the closest distance between the IRS~13E3 cluster and Sgr~A$^{\star}$ reported by Tsuboi et al.~\cite{Tsuboi1}. This distance sets the area where we average the expected ionization rate, as defined in Eq.~\eqref{eq:xiavg}.

\section{Results}
\label{sec:result}

\begin{figure*}[t]
    \centering
    \includegraphics[width=0.49\textwidth]{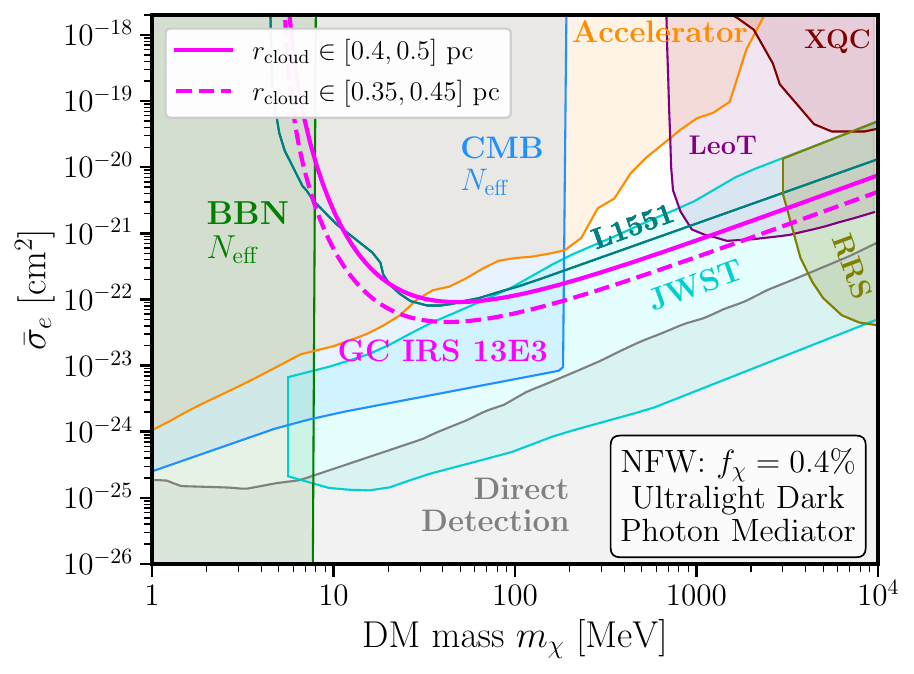}
    \hfill
    \includegraphics[width=0.49\textwidth]{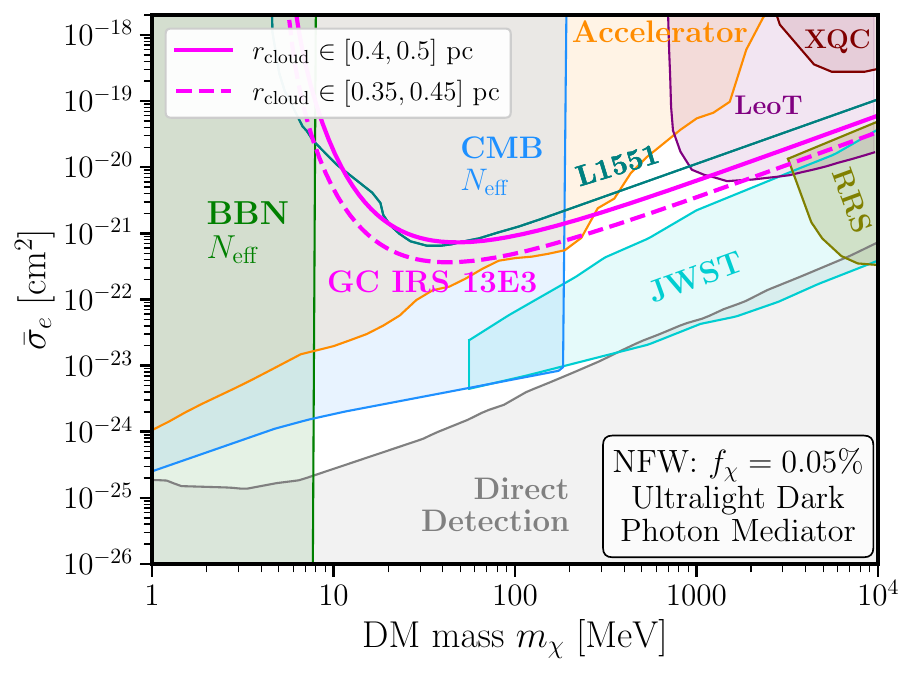}
    \caption{Constraints on $\bar{\sigma}_e$ for an ultralight dark photon mediator, $F_{\rm DM}(q) = (\alpha m_e/q)^2$, assuming an NFW profile, with \mbox{$f_\chi = 0.4\%$} ({\bf left}) and $f_\chi = 0.05\%$ ({\bf right}). The solid and dashed magenta curves show the CMZ constraints from GC~IRS~13E3 for $r_{\rm cloud} \in [0.4, 0.5]$~pc and $r_{\rm cloud} \in [0.35, 0.45]$~pc, respectively. Shaded regions are excluded by other probes: the molecular cloud L1551 (teal)~\cite{Prabhu:2022dtm}, JWST (cyan)~\cite{Du:2024afd}, direct detection (gray)~\cite{Emken:2019tni}, accelerators (orange)~\cite{Prinz:1998ua, Magill:2018tbb, ArgoNeuT:2019ckq, Plestid:2020kdm}, BBN and CMB~\cite{Creque-Sarbinowski:2019mcm, Munoz:2018pzp} $N_{\rm eff}$ (green and blue), Leo~T (purple)~\cite{Wadekar:2019mpc}, XQC (maroon)~\cite{Mahdawi:2018euy}, and RRS (olive)~\cite{Rich:1987st}.}
    \label{fig:constraint_light}
\end{figure*}

\begin{figure}[t]
    \centering
    \includegraphics[width=\columnwidth]{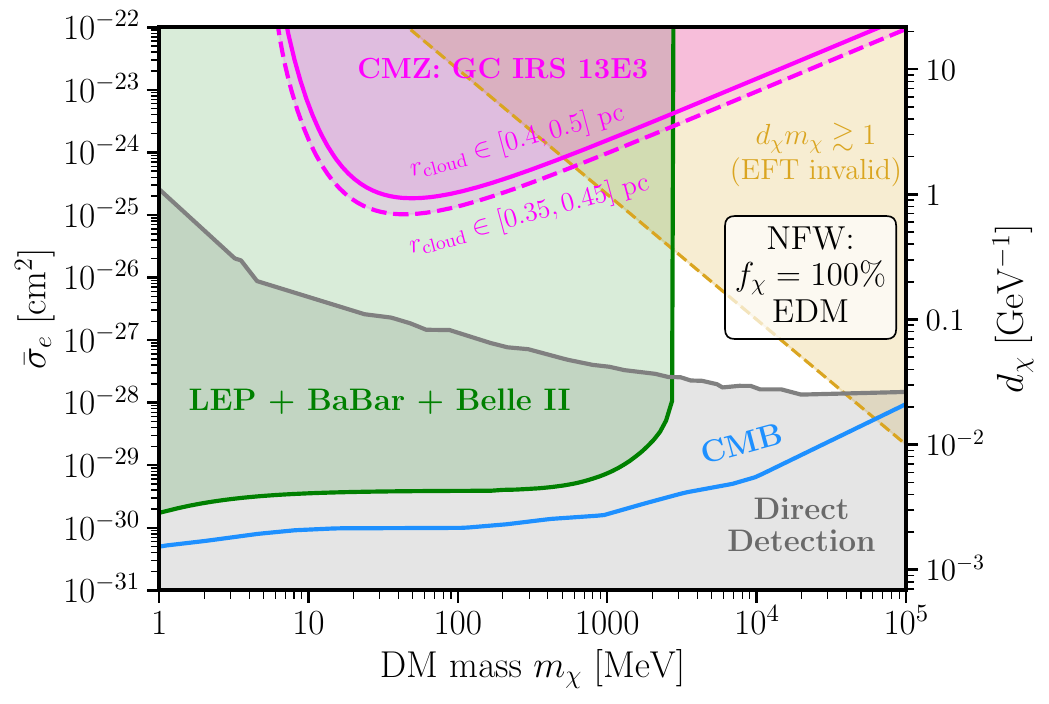}
    \caption{ Constraints on $\bar{\sigma}_e$ for DM with an electric dipole moment, with $f_\chi = 100\%$ and an NFW profile. The solid and dashed magenta curves show the CMZ constraints from GC~IRS~13E3 for $r_{\rm cloud} \in [0.4, 0.5]$~pc and $r_{\rm cloud} \in [0.35, 0.45]$~pc, respectively. Shaded regions are excluded by LEP, BaBar, and Belle~II (green)~\cite{Chu:2018qrm, Ibarra:2024mpq, Fortin:2011hv}, the CMB (blue)~\cite{Xu:2018efh}, and direct detection (gray)~\cite{Emken:2019tni}. In the golden region, $d_\chi m_\chi \gtrsim 1$ and the effective description of the EDM is not valid, see Eq.~\eqref{eq:EFT}. The right axis shows $d_\chi$ assuming $\mu_{\chi e} \simeq m_e$, which holds for $m_\chi \gg m_e$.
    }
    \label{fig:EDM}
\end{figure}

We derive constraints on the reference DM-electron cross section $\bar{\sigma}_e$ by requiring the DM-induced ionization rate not to exceed the observed rate, $\bar{\xi}_{\rm DM} \leq \zeta_{\rm obs}$. The regions above the magenta curves in Figs.~\ref{fig:constraint_heavy}, \ref{fig:constraint_light}, and~\ref{fig:EDM} are therefore excluded. Unless stated otherwise, we adopt an NFW density profile and show two positions for the molecular gas, $r_{\rm cloud} \in [0.35, 0.45]$~pc and $r_{\rm cloud} \in [0.4, 0.5]$~pc. Since Ref.~\cite{Tsuboi1} gives $r \gtrsim 0.4$~pc as a lower limit on the cloud position, the $[0.35, 0.45]$~pc interval places the cloud center at this lower limit, where the DM density is largest. Our limits therefore correspond to the most favorable position allowed by current observations. The $[0.4, 0.5]$~pc interval shows how the limits change when the cloud moves outward, and a cloud near 1~pc would weaken them by about an order of magnitude (Fig.~\ref{fig:Ionrate}).

The common shape of the bounds follows directly from the ionization kinematics. At low DM masses the available kinetic energy becomes insufficient to efficiently ionize H$_2$, causing the constraint to rapidly weaken below ${\cal O}(10)$~MeV. At larger masses the DM number density decreases as $\rho_\chi/m_\chi$, producing the gradual weakening of the bounds above the minimum at masses of a few tens of MeV. For a subdominant component, $\rho_\chi=f_\chi\times\rho_{\rm DM}$, so that the bound scales approximately as $\bar{\sigma}_e^{\rm lim}\propto f_\chi^{-1}$ when attenuation effects are negligible, down to $f_\chi \sim 10^{-6}$ for our fiducial gas column, below which the DM energy flux cannot sustain the observed ionization rate (see Appendix~\ref{appen:atten}).

Fig.~\ref{fig:constraint_heavy} shows the constraints for heavy mediators in the strongly interacting regime, at large $\bar{\sigma}_e$, where the CMZ measurement is most useful. For the heavy dark photon with $f_\chi = 100\%$ (left panel), the constraint is strongest at $m_\chi \sim 10$ to $100$~MeV and lies below the CR bound over a large part of the mass range. The CR bound requires that CR electrons, which scatter with DM as they propagate through the Galaxy, do not lose enough energy to distort the observed electron spectrum~\cite{Cappiello:2018hsu}. The CMZ probes cross sections well above the upper reach of terrestrial direct detection experiments, where repeated scattering in the atmosphere and overburden stops the halo DM before it reaches underground detectors. This is the main advantage of using molecular gas as the target itself, rather than detecting DM after it propagates to the Earth. Above $m_\chi \sim 100$~MeV, however, the CMZ curve lies above the attenuation threshold. There, DM-proton scattering both slows the DM and ionizes H$_2$ through the Migdal effect~\cite{Blanco:2023bgz}, so our calculation, which includes only DM-electron scattering, is no longer reliable (Appendix~\ref{appen:atten}). The CMB bound~\cite{Xu:2018efh} also lies four to six orders of magnitude below the CMZ curve. The CMZ therefore matters most for a subdominant component, which acts cosmologically as a small perturbation to the baryons.

For a subdominant dark photon, the CMZ curve rises as $f_\chi^{-1}$ and lies above the attenuation threshold at all masses, so our calculation does not apply to this model. The right panel of Fig.~\ref{fig:constraint_heavy} instead shows the leptophilic interaction with $f_\chi = 0.01\%$. At this fraction, CMB constraints on a subcomponent that is tightly coupled to baryons lose their power~\cite{Boddy:2018wzy, Prabhu:2022dtm}, and the CMZ constraint lies below the CR curve for $m_\chi \gtrsim 12$~MeV, which shows the gain from the high DM density near the GC. Without proton scattering, however, DM loses little energy in the Earth, and the direct detection region extends to $\bar{\sigma}_e \sim 10^{-18}$~cm$^2$ at $m_\chi = 100$~MeV. The CMZ region therefore lies inside the regions already excluded by direct detection and CRs. The CMZ and CR bounds both scale as $f_\chi^{-1}$, while the direct detection ceiling does not depend on $f_\chi$, so the CR bound would move above the direct detection ceiling only for $f_\chi \lesssim 10^{-7}$. At such small fractions, however, the kinetic energy flux of DM cannot sustain the observed ionization rate for our fiducial column, $N_{\rm H} \simeq 6\times10^{22}$~cm$^{-2}$, which requires $f_\chi \gtrsim 10^{-6}$ [Eq.~\eqref{eq:energy_budget}]. This minimum fraction scales with the column, which is not measured for IRS~13E3. For our fiducial column, the CMZ therefore does not probe new parameter space for the leptophilic interaction.

All our constraints also satisfy partial-wave unitarity, which limits the elastic cross section to \mbox{$\sigma_\ell \leq 4\pi(2\ell+1)/k^2$}, where $k = \mu_{\chi e} v$ is the center-of-mass momentum~\cite{Griest:1989wd, Digman:2019wdm}. For $v \simeq 300$~km/s and $m_\chi \gg m_e$, the $s$-wave bound gives \mbox{$\bar{\sigma}_e \lesssim 2\times10^{-14}$~cm$^2$}, independent of $m_\chi$ above 10~MeV. This value lies more than an order of magnitude above the largest cross sections in Fig.~\ref{fig:constraint_heavy}, so unitarity does not limit the parameter space we probe.

Fig.~\ref{fig:constraint_light} shows the constraints for the ultralight mediator, with $f_\chi = 0.4\%$ in the left panel and $f_\chi = 0.05\%$ in the right panel. We choose these DM fractions to match the benchmarks of the James Webb Space Telescope (JWST) analysis~\cite{Du:2024afd}, and to show how the different bounds respond to the DM fraction. The bounds from GC~IRS~13E3 move up, since the ionization rate scales with the DM density. The lower edge of the JWST bound moves up in the same way, but its upper edge is set by the attenuation of DM in the detector shielding, which depends only weakly on $f_\chi$~\cite{Du:2024afd}. At smaller $f_\chi$ the JWST band therefore narrows, and it survives only at larger masses, where this ceiling is higher. The form factor $|F_{\rm DM}|^2 \propto q^{-4}$ gives more weight to small momentum transfers, but the overall shape of the CMZ limit follows the same pattern as for the heavy mediator: the ionization threshold sets the low-mass edge, and the falling DM number density weakens the limit at large mass. At $m_\chi \lesssim 10$~MeV, the constraint for $f_\chi = 0.05\%$ lies above the attenuation threshold (Appendix~\ref{appen:atten}) and is not reliable there, but the BBN and CMB $N_{\rm eff}$ bounds already cover this region.

We also compare the GC~IRS~13E3 constraints with other bounds: the molecular cloud L1551~\cite{Prabhu:2022dtm}, and the direct detection ceiling~\cite{Emken:2019tni}, which combines the XENON10~\cite{Essig:2017kqs, XENON10:2011prx}, XENON100~\cite{XENON:2016jmt}, SENSEI~\cite{Crisler:2018gci, SENSEI:2019ibb}, DarkSide-50~\cite{DarkSide:2018ppu}, and CDMS-HVeV constraints~\cite{SuperCDMS:2018mne}. Accelerator bounds come from beam-dump~\cite{Prinz:1998ua} and neutrino experiments~\cite{Magill:2018tbb, ArgoNeuT:2019ckq, Plestid:2020kdm}. The figures further show the BBN and CMB $N_{\rm eff}$ bounds~\cite{Creque-Sarbinowski:2019mcm, Munoz:2018pzp}, the X-ray Quantum Calorimeter (XQC)~\cite{Mahdawi:2018euy}, the Rich-Rocchia-Spiro balloon experiment (RRS)~\cite{Rich:1987st}, and Leo~T gas heating~\cite{Wadekar:2019mpc}.

L1551 gives the most direct comparison, since its constraint comes from the same observable, the DM-induced ionization of H$_2$. The GC~IRS~13E3 curves lie below the L1551 constraint over most of the mass range, so the improvement comes from the choice of target rather than from the detection mechanism. Two factors compete. The NFW density at 0.4~pc is about $3\times10^4$ times the local value of 0.4~GeV\,cm$^{-3}$, while $\zeta_{\rm obs}$ in the CMZ is $10^2$ to $10^3$ times larger than the rate of $10^{-18}$ to $10^{-17}$~s$^{-1}$ in L1551~\cite{Prabhu:2022dtm}. Together they give a gain of about 30 to 300, consistent with the factor of about 50 at 100~MeV in Fig.~\ref{fig:constraint_light}. The velocity distribution near Sgr~A$^{\star}$ changes this ratio only by a factor of order unity, through the fraction of DM above $v_{\rm min}$ and through $\eta \propto 1/v$.

The comparison with JWST is different, since JWST searches for electron recoils in a detector in space~\cite{Du:2024afd}. For $f_\chi = 0.4\%$, the CMZ curve passes through the region that JWST already excludes. For $f_\chi = 0.05\%$, the CMZ curve lies above the upper edge of the JWST band over much of the sub-GeV mass range. At small $f_\chi$, the CMZ therefore reaches larger cross sections than JWST, since no shielding separates the DM from the molecular gas, which acts as the detector.

The remaining bounds probe different physics. Accelerator searches produce DM through its coupling to electric charge~\cite{Prinz:1998ua, Magill:2018tbb, ArgoNeuT:2019ckq, Plestid:2020kdm}, so they do not depend on the local DM density. XQC~\cite{Mahdawi:2018euy} and RRS~\cite{Rich:1987st} detect DM scattering above most of the atmosphere, and Leo~T constrains the heating of its gas by DM~\cite{Wadekar:2019mpc}. The BBN and CMB $N_{\rm eff}$ bounds instead come from the effect of DM on the thermal history of the early Universe~\cite{Creque-Sarbinowski:2019mcm, Munoz:2018pzp}. They assume that $\chi$ reaches thermal equilibrium with the SM plasma before BBN, or before recombination for the CMB bound~\cite{Creque-Sarbinowski:2019mcm}, so unlike the ionization bound derived here, they depend on the thermal history of the dark sector.

Finally, Fig.~\ref{fig:EDM} shows the constraints for the EDM interaction, where $|F_{\rm DM}|^2 \propto q^{-2}$ lies between the heavy and ultralight cases. The CMZ constraint is strongest at a few tens of MeV, where it reaches $\bar{\sigma}_e \sim 10^{-25}$~cm$^2$. For $f_\chi = 100\%$, the CMB bound lies several orders of magnitude below the CMZ curve, and LEP, BaBar, and Belle~II exclude the CMZ region for $m_\chi \lesssim 3$~GeV. At larger masses, the CMZ curve lies in the region $d_\chi m_\chi \gtrsim 1$, where an elementary EDM is not a valid description. There, DM can instead be a composite state whose dipole comes from its internal structure, as for atoms and molecules. Our calculation still applies to such a state if its size is small compared to the inverse momentum transfer in the cloud, $1/q \sim 5\times10^{-9}$~cm, although a composite state can also have other interactions, such as a charge radius or a polarizability, that we do not include. The collider bounds do not depend on $f_\chi$, while the CMB bound is expected to weaken for a subdominant component, as for the dark photon. At $m_\chi \gtrsim 3$~GeV, however, the CMZ curve lies only a factor of a few below the attenuation threshold (Appendix~\ref{appen:atten}), so for a subdominant component it moves above the threshold and our calculation no longer applies.

Together, these results show where the CMZ adds to existing probes. The gain comes from the high DM density near Sgr~A$^{\star}$, which improves on previous molecular cloud bounds with the same detection mechanism. Among the bounds we show, the CMZ gives the leading constraint for the ultralight mediator with $f_\chi = 0.05\%$, above the JWST band over much of the sub-GeV mass range. For the heavy dark photon, attenuation in the cloud limits the constraint to $m_\chi \lesssim 100$~MeV at \mbox{$f_\chi = 100\%$}, where the CMB bound is stronger, and for the leptophilic interaction the constraint lies inside existing bounds. For the EDM, the CMZ constrains only the region where an elementary dipole is not valid, at $f_\chi = 100\%$.

\section{Conclusion and Outlook}
\label{sec:conclusion}

We have studied the ionization of molecular hydrogen in the CMZ as a probe of DM-electron scattering. The high DM density near the GC, together with measurements of the ionization rate through H$_3^+$, makes the CMZ a useful target for strongly interacting DM, which underground searches cannot reach because of the atmosphere and the Earth's crust. We considered heavy and ultralight dark photon mediators, a leptophilic contact interaction, and an EDM interaction, and we derived constraints by requiring the DM-induced ionization rate in the gas of the GC~IRS~13E3 cluster~\cite{Tsuboi1} not to exceed the observed CMZ ionization rate.

\begin{figure*}[t]
    \centering
    \includegraphics[width=2\columnwidth]{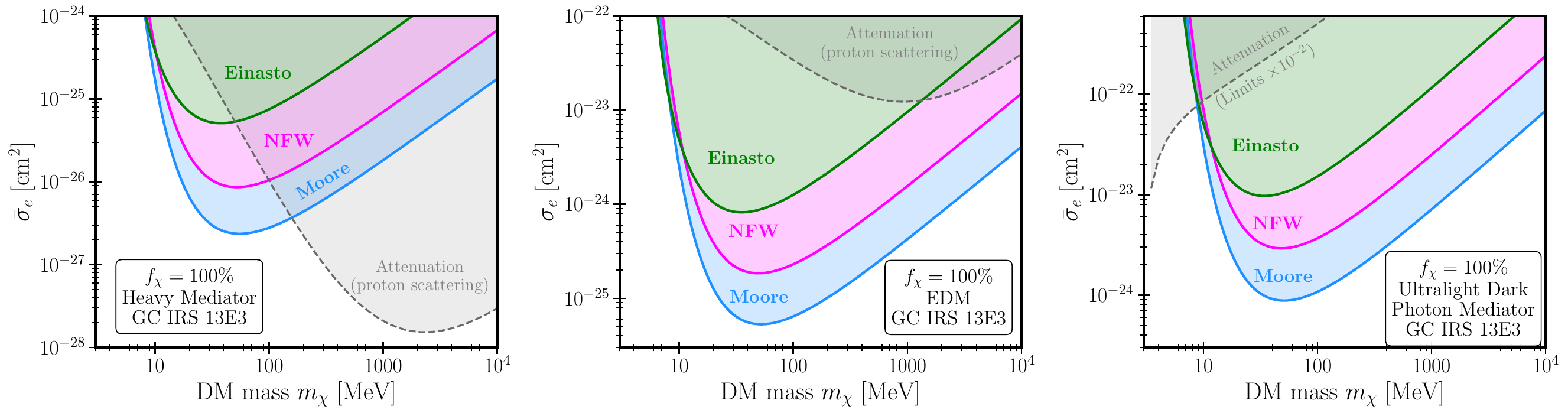}
    \caption{Constraints on the reference DM-electron cross section for the Galactic Center IRS~13E3 cluster ($r_{\rm cloud}\in[0.4,0.5]$~pc) from the CMZ ionization observation, using different DM density profiles: Einasto (green), NFW (magenta), and Moore (blue). The three panels show the heavy mediator with $F_{\rm DM}=1$ ({\bf left}), the EDM interaction with $F_{\rm DM}=\alpha m_e/q$ ({\bf middle}), and the ultralight dark photon mediator with $F_{\rm DM}=(\alpha m_e/q)^{2}$ ({\bf right}). The dashed gray curves show the attenuation threshold from DM-proton scattering in the cloud, assuming DM in the tail of the velocity distribution, at the largest speed allowed at $r = 0.4$~pc; above them, no DM can ionize H$_2$ at the cloud center, and our constraints are not reliable (Appendix~\ref{appen:atten}). In the left panel, the threshold applies to the heavy dark photon, and the leptophilic interaction has no proton attenuation. In the right panel, we multiply the threshold by $10^{-2}$ to show it in the plotted range, and the actual threshold lies far above the constraints and does not affect them.}
    \label{fig:Uncertainty}
\end{figure*}

A potential issue in the strongly interacting regime is that, for the dark photon and EDM interactions, DM also scatters with protons in the cloud before it reaches the molecular gas we consider. This scattering slows the DM and ionizes H$_2$ through the Migdal effect~\cite{Blanco:2022pkt, Blanco:2023bgz}. We computed the attenuation threshold above which proton scattering stops the DM before it reaches the cloud center, and we use it to determine which of our constraints are reliable (Fig.~\ref{fig:Uncertainty}, Appendix~\ref{appen:atten}). Determining whether proton scattering reduces or increases the ionization rate requires a full treatment including the Migdal effect as well, which we leave for future work.

For the heavy dark photon with $f_\chi = 100\%$, the CMZ constrains large cross sections at $m_\chi \lesssim 100$~MeV, although the CMB bound is stronger there. At larger masses, and at all masses for a subdominant component, the CMZ curve lies above the attenuation threshold, where DM-proton scattering makes our calculation unreliable in the non-leptophilic scenario. The leptophilic interaction avoids proton scattering, but for $f_\chi = 0.01\%$ its constraint lies inside the regions excluded by direct detection and CRs, and, for our fiducial gas column, the energy budget of the DM flux prevents it from probing new parameter space at smaller fractions. For the ultralight mediator, the GC~IRS~13E3 constraint improves on the L1551 limit~\cite{Prabhu:2022dtm} by about a factor of 50 at 100~MeV, which shows the gain from the high DM density in the inner Galaxy. JWST~\cite{Du:2024afd} gives a stronger bound for $f_\chi = 0.4\%$, while for $f_\chi = 0.05\%$ the CMZ reaches larger cross sections than JWST over much of the sub-GeV mass range. For the EDM, collider and CMB bounds are stronger wherever an elementary dipole is valid, and the CMZ constrains the region $d_\chi m_\chi \gtrsim 1$ for $f_\chi = 100\%$, where DM can be a composite state.

The main astrophysical uncertainty is the DM distribution in the central parsec. Our limits depend on both the cloud position and the inner DM profile, while the gravitational potential of Sgr~A$^\star$ and the nuclear star cluster shape the DM velocity distribution. Relative to NFW, the Einasto profile weakens the limits by up to an order of magnitude, while the Moore profile strengthens them by a factor of a few, for all interactions (Fig.~\ref{fig:Uncertainty}, Appendix~\ref{appen:uncertainty}). Better measurements of the three-dimensional position and the ionization rate of molecular gas close to the GC, together with a better model of the inner DM phase-space distribution, would directly improve the reach of this method.

More generally, molecular gas in the central Galaxy acts as a detector for DM interactions in a regime that is difficult to reach from the Earth. A measurement of the ionization rate at the position of GC~IRS~13E3 would remove the need to use the CMZ average, although it would also include the ionization from young stars in the cluster. The IRS~13 cluster contains many dusty sources that are bright in the L~band~\cite{Peissker2023}, so H$_3^+$ absorption toward these sources, observed with high-resolution infrared spectrographs such as CRIRES+~\cite{Dorn2023}, could provide such a measurement. Together with a better determination of the three-dimensional position of the gas, it would make the CMZ a sharper probe of subdominant, strongly interacting DM.

\section*{Acknowledgements}
\vspace*{-0.3cm}

We thank Phan Vo Hong Minh for discussions on CMZ observations, Chris Cappiello for discussions on strongly interacting dark matter, and Nick Rodd for discussions on the Milky Way halo model. TTQN is supported by the Swedish Research Council under contract 2022-04283, and by two grants from the Royal Swedish Academy of Sciences~(KVA): PH2025-0073 (Physics) and AST2025-0048 (Astronomy and Space Science). TTQN acknowledges the support from EDUCATE Excellence Centre funded by the Swedish Research Council through grant Dnr~2022-06627. PDL is supported by Ramón y Cajal RYC2024-048445-I grant, which is funded by MCIU/AEI/10.13039/501100011033 and FSE+. The work of PDL is also supported by the grants PID2021-125331NB-I00 and CEX2020-001007-S, both funded by MCIN/AEI/10.13039/501100011033 and by ``ERDF A way of making Europe''. PDL also acknowledges the MultiDark Network, ref. RED2022-134411-T. SB is supported by the STFC under grant ST/X000753/1. CB is supported by The Charles E. Kaufman Foundation of The Pittsburgh Foundation. Parts of this work were performed using computing resources provided by the National Academic Infrastructure for Supercomputing in Sweden (NAISS) under Project 2025/5-729, which is partially funded by the Swedish Research Council through Grant 2022-06725. This research also utilized the Sunrise HPC facility supported by the Technical Division at the Department of Physics, Stockholm University.
\\

\noindent {\bf AI Usage Statement.} Claude Code was used to assist with visually generating the plots as well as for grammar and typographical checks. AI tools were not used to generate scientific results, draw conclusions or to write any sections of the paper. All analysis, code, and the final text were produced, checked, and verified by the authors, who take full responsibility for the content of this work.

\appendix

\section{Uncertainties from the DM density profile}
\label{appen:uncertainty}

The main astrophysical uncertainty in our constraints is the DM distribution in the central parsec. We compute the limits for three spherically symmetric profiles with different inner slopes, using the parameters of Ref.~\cite{Cirelli:2010xx}:
\begin{align}
\text{NFW:} \quad & \rho_{\rm DM}(r) = \rho_s \frac{r_s}{r}\left(1 + \frac{r}{r_s}\right)^{-2}, \label{eq:NFW}\\
\text{Einasto:} \quad & \rho_{\rm DM}(r) = \rho_s \exp\left\{-\frac{2}{\alpha}\left[\left(\frac{r}{r_s}\right)^{\alpha} - 1\right]\right\}, \label{eq:Einasto}\\
\text{Moore:} \quad & \rho_{\rm DM}(r) = \rho_s \left(\frac{r_s}{r}\right)^{1.16}\left(1 + \frac{r}{r_s}\right)^{-1.84}, \label{eq:Moore}
\end{align}
where $r$ is the distance from the GC. The NFW profile is the standard benchmark from $N$-body simulations~\cite{Navarro:1995iw, Navarro:1996gj}, the Einasto profile fits more recent simulations and becomes shallower toward the center~\cite{Graham:2005xx}, and the Moore profile~\cite{Diemand:2004wh} is steeper than NFW. All three profiles give a local density of $0.3$~GeV\,cm$^{-3}$ at $r_\odot = 8.33$~kpc. Table~\ref{tab:profiles} lists the parameters and the density at $r = 0.4$~pc.

\begin{table}[t]
\vspace{0.3cm}
\centering
\renewcommand{\arraystretch}{1.3}
\setlength{\tabcolsep}{5pt}
\begin{tabular}{lcccc}
\hline\hline
Profile & $\alpha$ & $r_s$ & $\rho_s$ & $\rho_{\rm DM}(0.4\,{\rm pc})$ \\
 & & [kpc] & [GeV\,cm$^{-3}$] & [GeV\,cm$^{-3}$] \\
\hline
NFW & & 24.42 & 0.184 & $1.1\times10^4$ \\
Einasto & 0.17 & 28.44 & 0.033 & $7.3\times10^2$ \\
Moore & & 30.28 & 0.105 & $4.8\times10^4$ \\
\hline\hline
\end{tabular}
\caption{Parameters of the DM density profiles~\cite{Cirelli:2010xx} and the DM density at the position of GC~IRS~13E3.}
\label{tab:profiles}
\end{table}

For each profile, we compute the velocity distribution with the same Milky Way mass model as in Sec.~\ref{sec:Ionization_rate}. Fig.~\ref{fig:Uncertainty} shows the CMZ constraints for the three profiles and the three interactions, with $f_\chi = 100\%$. Relative to NFW, the Einasto density at 0.4~pc is 15 times smaller, and the Moore density is 4 times larger. The limits do not scale exactly with the density, since the profile also changes the velocity distribution. At $r \sim 0.4$~pc, Sgr~A$^\star$ dominates the potential, so $v_c \simeq 215$~km/s for all profiles, but the Jeans equation gives $\sigma_v^2 \simeq v_c^2/(1+\gamma)$, where $\gamma$ is the local logarithmic slope of the density. The Einasto profile is shallow at this radius, $\gamma \simeq 0.3$, which gives $\sigma_v \simeq 190$~km/s, compared with about 150~km/s for NFW and Moore. The larger dispersion puts more DM above $v_{\rm min}$ and partly compensates the lower Einasto density, most strongly at low masses near the ionization threshold. As a result, the Einasto profile weakens the limits by a factor of 5 to 15 relative to NFW, depending on the DM mass, while the Moore profile strengthens them by about a factor of 4. A generalized NFW profile with a steeper inner slope would strengthen the constraints further.

\section{Attenuation of the DM flux in the cloud}
\label{appen:atten}

For the dark photon and EDM interactions, DM scatters with the protons in the cloud and slows down (Sec.~\ref{sec:DMmodel}). A DM particle can ionize H$_2$ only if its kinetic energy exceeds the ionization energy of Eq.~\eqref{eq:Eion}, $m_\chi v^2/2 > E_{\rm ion}$, since this is the largest energy it can deposit. Here we estimate the cross section above which the DM loses this ability before it reaches the cloud center.

We describe the elastic scattering of DM on a hydrogen atom, in which the atom stays in its ground state. In the nonrelativistic limit, ${\rm d}\sigma/{\rm d}q^2$ does not depend on the target mass, so the cross section on the atom follows from Eq.~\eqref{eq:dsigma},
\begin{equation}
    \frac{{\rm d}\sigma_{\rm H}}{{\rm d}q^2} = \frac{\bar{\sigma}_e}{4\mu_{\chi e}^2 v^2}\,|F_{\rm DM}(q)|^2\,|Z_{\rm eff}(q)|^2,
    \label{eq:dsigma_H}
\end{equation}
where $Z_{\rm eff}(q)$ is the effective charge of the atom at momentum transfer $q$. The amplitude is the sum of the amplitudes on the proton and on the electron cloud. The proton is pointlike at these momenta, and the electron cloud gives the hydrogen form factor
\begin{equation}
    F_{\rm H}(q) = \left[1 + \left(\frac{q a_0}{2}\right)^2\right]^{-2},
    \label{eq:FH_elastic}
\end{equation}
with $a_0$ the Bohr radius defined below Eq.~\eqref{eq:fH}. Since DM couples to electric charge, the proton and the electron enter with opposite signs, and $Z_{\rm eff}(q) = 1 - F_{\rm H}(q)$. The neutral atom screens the proton at $q \lesssim 1/a_0$, while at $q \gg 1/a_0$ the DM scatters on the bare proton and $Z_{\rm eff} \to 1$.

Each scattering gives the atom a recoil energy $q^2/(2m_{\rm H})$, where $m_{\rm H}$ is the hydrogen mass, with \mbox{$0 \leq q \leq q_{\rm max} = 2\mu_{\chi \rm H} v$} and $\mu_{\chi \rm H}$ the DM-hydrogen reduced mass. The DM therefore loses energy along its path at the rate
\begin{align}
    \frac{{\rm d}E}{{\rm d}s} &= -n_{\rm H}\,\bar{\sigma}_e\,\mathcal{S}(v), \nonumber\\
    \mathcal{S}(v) &= \int_0^{q_{\rm max}^2} {\rm d}q^2\, \frac{|F_{\rm DM}(q)|^2 |Z_{\rm eff}(q)|^2}{4\mu_{\chi e}^2 v^2}\,\frac{q^2}{2m_{\rm H}}.
    \label{eq:stopping}
\end{align}
We follow a DM particle with initial kinetic energy \mbox{$E_0 = m_\chi v_0^2/2$} along a straight path of length $s$ and treat the slowing down as continuous. The DM kinetic energy falls to $E_{\rm ion}$ at the end of the path when
\begin{equation}
    \bar{\sigma}_e^{\rm att} = \frac{1}{n_{\rm H}\, s}\int_{E_{\rm ion}}^{E_0} \frac{{\rm d}E}{\mathcal{S}(E)},
    \label{eq:sigma_att}
\end{equation}
with $v = (2E/m_\chi)^{1/2}$ inside $\mathcal{S}$. We take \mbox{$n_{\rm H} = 4\times10^5$~cm$^{-3}$} (Sec.~\ref{sec:observation}) and $s = \Delta r/2 = 0.05$~pc, the distance from the cloud surface to its center, which gives a column density $N_{\rm H} = n_{\rm H}s \simeq 6.2\times10^{22}$~cm$^{-2}$. For $v_0$ we take the tail of the velocity distribution, the largest DM speed in the cloud frame at $r = 0.4$~pc, $v_0 = v_e + v_c \simeq 955$~km/s. Above $\bar{\sigma}_e^{\rm att}$, even the fastest DM stops before it reaches the cloud center.

For the heavy dark photon, large momentum transfers dominate the energy loss, so $Z_{\rm eff} \simeq 1$ and Eq.~\eqref{eq:stopping} gives ${\rm d}E/{\rm d}s = -\kappa E$, with
\begin{equation}
    \kappa = \frac{2 n_{\rm H}\bar{\sigma}_e\,\mu_{\chi \rm H}^4}{m_\chi m_{\rm H}\,\mu_{\chi e}^2}.
    \label{eq:kappa}
\end{equation}
Every DM particle therefore loses the same fraction of its energy along its path, $E(s) = E_0\,e^{-\kappa s}$, whatever its initial speed, and Eq.~\eqref{eq:sigma_att} gives
\begin{equation}
    \bar{\sigma}_e^{\rm att} \simeq \frac{m_\chi m_{\rm H}\,\mu_{\chi e}^2}{2N_{\rm H}\,\mu_{\chi \rm H}^4}\,\ln\frac{E_0}{E_{\rm ion}}.
    \label{eq:sigma_att_heavy}
\end{equation}
The threshold depends on $v_0$ only through the logarithm and is smallest at \mbox{$m_\chi = 3m_{\rm H}$}, where \mbox{$\bar{\sigma}_e^{\rm att} \simeq 2\times10^{-28}$~cm$^2$}. At $m_\chi = 50$~MeV, \mbox{$\bar{\sigma}_e^{\rm att} \simeq 5.4\times10^{-26}$~cm$^2$} for $v_0 = 955$~km/s and $2.9\times10^{-26}$~cm$^2$ for $v_0 = 500$~km/s. Below the threshold the DM still slows down: at the NFW constraint for $m_\chi = 50$~MeV, $\bar{\sigma}_e \simeq 8\times10^{-27}$~cm$^2$, the DM reaches the cloud center with about two thirds of its initial energy. Constraints close to the threshold are therefore weakened by attenuation, and only those well below it are unaffected. For the EDM and the ultralight dark photon, $F_{\rm DM}$ gives more weight to small $q$, where the atom screens the proton and each scattering transfers little energy. Their thresholds are therefore much higher, $\bar{\sigma}_e^{\rm att} \simeq 3.6\times10^{-23}$~cm$^2$ and $4.7\times10^{-20}$~cm$^2$ at $m_\chi = 100$~MeV, respectively, and grow roughly as $v_0^2$. Notably, we also compute the attenuation from DM scattering on the electron cloud alone, with $Z_{\rm eff}(q) = F_{\rm H}(q)$ in Eq.~\eqref{eq:dsigma_H}. The resulting threshold lies orders of magnitude above the one from proton scattering and does not affect the constraints in this work.

Fig.~\ref{fig:Uncertainty} compares $\bar{\sigma}_e^{\rm att}$ with the constraints for the three DM profiles. For the heavy dark photon, the constraints cross the threshold at $m_\chi \simeq 50$, $100$, and $200$~MeV for the Einasto, NFW, and Moore profiles, respectively. Above these masses even the fastest DM stops before the cloud center, and near them attenuation weakens the constraints. For the EDM, the NFW and Moore constraints lie below the threshold over the whole mass range, while the Einasto constraint crosses it at $m_\chi \simeq 1$~GeV. For the ultralight dark photon, the threshold lies more than two orders of magnitude above all the constraints.

Eq.~\eqref{eq:sigma_att} includes only elastic scattering. Energy lost to excitation and ionization of H$_2$ is negligible for the heavy and ultralight dark photon, and about 25\% of the elastic loss for the EDM, so we slightly overestimate the EDM threshold. We also treat the gas as at rest, follow a straight path, and treat each hydrogen atom in H$_2$ as independent, which changes the threshold by a factor of order unity.

The kinetic energy flux of DM also limits the ionization rate that DM can sustain in the cloud, for any interaction. Each ionization costs at least $E_{\rm ion}$, so sustaining $\zeta_{\rm obs}$ over the column from the cloud surface to its center requires an energy flux $N_{\rm H}\times \zeta_{\rm obs}\times E_{\rm ion}/2$. The DM supplies at most $f_\chi\times \rho_{\rm DM}\times v^3/2$, which gives the necessary condition
\begin{align}
    f_\chi \gtrsim\ & 1.3\times10^{-6}
    \left(\frac{N_{\rm H}}{6.2\times10^{22}\,{\rm cm}^{-2}}\right)
    \left(\frac{\zeta_{\rm obs}}{2\times10^{-15}\,{\rm s}^{-1}}\right) \nonumber\\
    &\times \left(\frac{1.1\times10^4\,{\rm GeV\,cm}^{-3}}{\rho_{\rm DM}}\right)
    \left(\frac{500\,{\rm km/s}}{v}\right)^3.
    \label{eq:energy_budget}
\end{align}
This condition is optimistic, since the DM flux is not a beam and only part of its energy goes into ionization. Below this fraction, DM cannot produce the observed ionization rate for any cross section, and the $f_\chi^{-1}$ scaling of our constraints no longer holds.

\bibliography{ref}

\end{document}